\documentclass[twocolumn]{aastex63}

\received{Jan 28, 2022}
\revised{Mar 11, 2022}
\accepted{Mar 15, 2022}

\submitjournal{ApJS}

\shorttitle{Kepler Pixel Project}
\shortauthors{Forró et al.}

\graphicspath{{./}{figures/}}

\begin{document}

\title{Kepler Pixel Project – background RR Lyrae stars in the primary Kepler mission field of view}

\correspondingauthor{Adrienn Forró}
\email{forro.adrienn@csfk.org}
\author[0000-0001-9394-3531]{Adrienn Forró}
\affiliation{Konkoly Observatory, Research Centre for Astronomy and Earth Sciences (ELKH),\\ H-1121 Konkoly Thege Miklós út 15-17, Budapest, Hungary}
\affiliation{MTA CSFK Lendület Near-Field Cosmology Research Group, H-1121 Konkoly Thege Miklós út 15-17, Budapest, Hungary}
\affiliation{Eötvös Loránd University, Pázmány Péter sétány 1/A, Budapest, Hungary}

\author[0000-0002-3258-1909]{Róbert Szabó}
\affiliation{Konkoly Observatory, Research Centre for Astronomy and Earth Sciences (ELKH),\\ H-1121 Konkoly Thege Miklós út 15-17, Budapest, Hungary}
\affiliation{MTA CSFK Lendület Near-Field Cosmology Research Group, H-1121 Konkoly Thege Miklós út 15-17, Budapest, Hungary}
\affiliation{ELTE Eötvös Loránd University, Institute of Physics, Budapest, Hungary\\}

\author[0000-0002-8585-4544]{Attila Bódi}
\affiliation{Konkoly Observatory, Research Centre for Astronomy and Earth Sciences (ELKH),\\ H-1121 Konkoly Thege Miklós út 15-17, Budapest, Hungary}
\affiliation{MTA CSFK Lendület Near-Field Cosmology Research Group, H-1121 Konkoly Thege Miklós út 15-17, Budapest, Hungary}

\author[0000-0002-7947-9346]{Kornél Császár}
\affiliation{Eötvös Loránd University, Pázmány Péter sétány 1/A, Budapest, Hungary}
\affiliation{Konkoly Observatory, Research Centre for Astronomy and Earth Sciences (ELKH),\\ H-1121 Konkoly Thege Miklós út 15-17, Budapest, Hungary}

\begin{abstract}

 In this paper we describe
a project we initiated to investigate individual pixels in the downloaded $Kepler$ apertures in order to find objects in the background of the main targets with variable brightness. In the first paper of this series we discovered and investigated 547 short-period eclipsing binaries \citep{bienias2021}. Here we present the independent discovery of 26 new RR\,Lyrae stars in the \textit{Kepler} background pixels obtained during the primary mission, and provide continuous and precise photometry for these objects.
Twenty-one of these stars were already noted by Gaia or the Pan-STARRS survey. This new population of dominantly faint and distant RR\,Lyrae stars increases by 50\% and complements nicely the 52 already known main target RR\,Lyrae stars in the original \textit{Kepler} field. Despite their faintness, the four-year quasi--uninterrupted light curves of these stars allow an unprecedented view of these faint halo objects. We present an analysis of the light curves of the new RR\,Lyrae sample, verify their classification using Fourier parameters,  and discuss the properties of these newly found pulsating variable stars. Most notably, this is the first time that such faint RR\,Lyrae stars have been investigated with the help of a photometric data set with outstanding cadence and precision. Interestingly, these objects share the properties of their brighter siblings in terms of sub-class characteristics, additional mode content, and modulation occurrence rates.

\end{abstract}


\section{Introduction} \label{sec:intro}

The original {\it Kepler} mission provided unprecedented high-precision, quasi-uninterrupted, long photometric observations of nearly 200,000 stars \citep{borucki2010}. Its primary aim was to deliver reliable statistics of the frequency of Earth-like planets around solar-type stars via the transit method. In reality, {\it Kepler} achieved  much more than that, since the mission opened a new era in exoplanetary science, delivering a plethora of planets and planetary systems around a huge variety of stars challenging planet formation models. Among the notable discoveries we find the first multiply transiting planetary system \citep{holman2010}, planets orbiting binary stars \citep{doyle2011}, 
and the first transiting exoplanets in open clusters \citep{meibom2013} just to mention a few examples.

Similarly, along with CoRoT \citep{baglin2009}, MOST \citep{matthews2000}, and the BRITE Constellation \citep{weiss2014}, the {\it Kepler} mission revolutionized stellar astrophysics as well. 
Following the success of helioseismology, observational asteroseismology became an  established discipline using both solar-type stars \citep{chaplin2011} and red giants \citep{beck2011}, with important applications including the possibility to  distinguish between helium-core and hydrogen-shell burning on the giant branch  \citep{bedding2011} and internal angular momentum transport studies \citep{aerts2019}, as well as seismology in open clusters \citep{stello2011a, stello2011b, miglio2012}. Beyond solar-type stars and red giants, the detailed seismology of A--F-type main-sequence stars \citep{grigahcene2010, uytterhoeven2011}, high-mass stars  \citep{papics2014, papics2017}, white dwarfs \citep{bischoff2011}, and sub-dwarfs \citep{reed2011}
have become possible as well. \textit{Kepler} opened new avenues in the study of eclipsing binary stars as well \citep{prsa2011}, with over 3000 eclipsing binary stars listed in the Kepler Eclipsing Binary Catalog\footnote{\url{http://keplerebs.villanova.edu/}}.

\begin{figure*}
    \centering
    \includegraphics[scale=0.4]{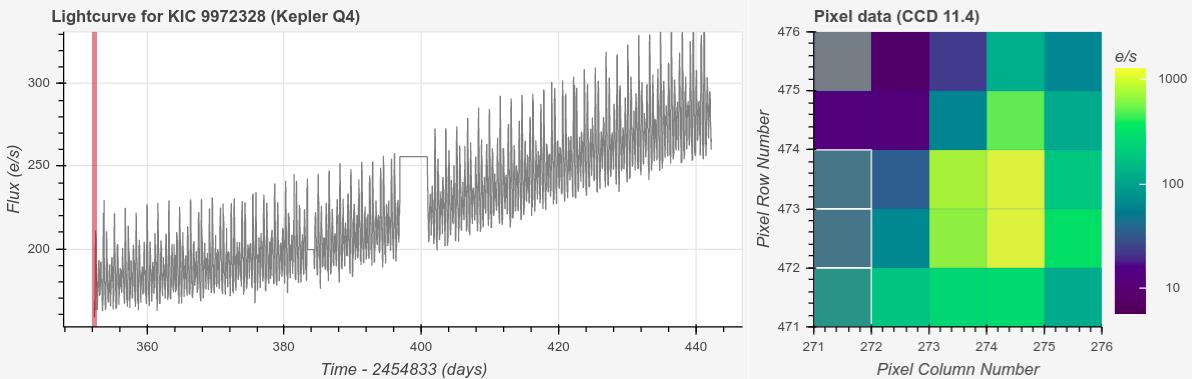}
    \caption{Light curve (left) and pixel selection (right) of a newly found RRd star on the Target Pixel File (TPF). The main target is centered on pixel (274,473), while the new variable star is at the lower left corner (three pixels outlined with white borders: (271,471), (271,472), and (271,473)). The plot was made by the {\tt Lightkurve} package {\bf \citep{Lightkurve}}.}
    \label{fig:pix_sel}
\end{figure*}

The Kepler Asteroseismic Science Consortium (KASC) was set
up to exploit the potential of \textit{Kepler} for solar-like oscillations as well as in all types of pulsations. KASC Working Group 13 is dedicated to the investigation of RR Lyrae stars observed by \textit{Kepler}. 

RR\,Lyrae stars are low-mass, high-amplitude pulsators on the horizontal branch. They belong to the old population, and because of their high and well-defined luminosity they are excellent tracers of galactic structures. These object primarily pulsate in low-order radial modes. RRab stars pulsate in the fundamental mode, RRc stars are first overtone pulsators, while RRd stars pulsate in both modes simultaneously \citep{catelansmith2015}. About half of the RRab stars and a significant fraction of the RRc stars show the famous Blazhko effect which is a conspicuous amplitude and/or period modulation, the  physical mechanism driving it is still far from being well-understood \citep{blazhko1907,kurtz2022}.

Out of 36 RRab stars that \textit{Kepler} observed in the original mission, 17 were found to be Blazhko stars \citep{benko2014, benko2015} based on the early data releases from the mission. 

{\it Kepler}-quality RR~Lyrae light curves enabled the discovery of new phenomena, like period doubling \citep{szabo2010}, additional modes \citep{moskalik2015}, resonances \citep{kollath2011,molnar2012} and the study of nonlinear dynamical phenomena in detail, like the complexity and multiplicity of the Blazhko modulation \citep{benko2010,benko2014}, as well as its temporal behavior \citep{guggenberger2012}. In addition, all RR~Lyrae light curves are unique and extremely rich in information which motivates the search for more representatives of this class, see for example the work of \cite{benko2019}. Therefore, any new addition to the inventory of RR\,Lyrae stars observed by \textit{Kepler} can be highly valuable. Moreover, as canonical RRd stars are quite rare, none has been found in the original {\it Kepler} field so far.

Part of our motivation was that we found faint, background RR~Lyrae stars serendipitously in the original {\it Kepler} data. Also, new dwarf novae and other transients (supernovae, quasar outbursts) were discovered in the aperture of other {\it Kepler} and K2 targets \citep{barclay2012,kato2013,kato2014,brown2015,ridden-harper2019, ridden-harper2020}, while solar-like oscillation signals were detected in some background {\it Kepler} pixels \citep{hon2018a}. \cite{colman2017} found foreground and background eclipsing binaries when analyzing oscillations of red giant stars among the {\it Kepler} main targets.
Recently, \citet{jackman2021} published the results of a systematic search for flare events in background objects in the apertures of $Kepler$ main targets observed in short (1-min) cadence.
 
In addition, synthetic galactic models, like TRILEGAL \citep{girardi2012} and Besan\c{c}on \citep{robin2003} show that the \textit{Kepler} RR Lyrae sample that was pre-selected by the KASC is seriously biased toward brighter magnitudes, thus faint, halo RR Lyrae stars are missing from our sample \citep{hanyecz2016}.

These findings motivated us to initiate a new program that we named the Kepler Pixel Project in order to explore new approaches to the {\it Kepler} data and to discover hitherto unknown pulsating stars and other time-variable objects. In this program, we examine individual pixels of the original {\it Kepler} mission to find interesting objects around the main {\it Kepler} targets. Specifically, we launched a project to find background, faint RR~Lyrae stars that are missing from the {\it Kepler} sample. Since the photometric quality is extremely high \citep{gilliland2010a}, and based on the examples presented above, we expect to find variable objects (especially large-amplitude variable stars) in those pixels that used to be considered as `background' of the main {\it Kepler} targets.

This paper is devoted to the discussion of the methods in general and to finding faint background RR Lyrae stars in particular. In Section~\ref{sec:methods} we present the data processing and variable star search methodology, in Section~\ref{sec:results} our results, in Section~\ref{sec:discussion} we discuss our main findings, and finally in  Section~\ref{sec:concl} we summarize our conclusions. In a series of companion papers we publish similar results on background eclipsing binaries (\citealt{bienias2021}; Forr\'o et al in prep.). We make the light curves of all the new variable stars publicly available\footnote{\url{https://www.konkoly.hu/KIK/data\_en.html}}. 

\section{Methods} \label{sec:methods}

\subsection{Kepler data and light curves}

During \textit{Kepler}'s original mission the spacecraft was observing one fixed field of view spanning an area of 105 deg$^2$ located at the northern celestial hemisphere containing parts of the constellations Cygnus, Lyra, and Draco. It was planned to be operated for 3.5 years, but it continued scientific observations for almost 4 years (from May 2009 March to May 2013), until the failure of its second reaction wheel. After a period of re-planning, the spacecraft was resurrected and started its second, Ecliptic K2 Mission \citep{howell2014}.
In this paper, however, we will exclusively use data gathered during the primary {\it Kepler} mission.

In order to
ensure optimal solar irradiation of the solar arrays, a 90 degree roll of the telescope was performed at the end of each quarter. After a short commissioning phase (Q0), the first quarter
lasted only for 33.5 d (Q1). The second quarter was the first complete
one (Q2). All subsequent quarters lasted roughly for 90–100 days, except the last one, Q17, which had a length of 30 days. Rotation of the spacecraft means that objects fell onto four different CCDs over the four quarters within each year. 

The \textit{Kepler} magnitude system (Kp) refers to the wide passband
(430–900 nm) transmission of the telescope and detector system.
Both long-cadence (LC, 29.4 min; \citealt{jenkins2010}), and short-cadence (SC, 58.9 s; \citealt{gilliland2010b}) observations are based
on the same 6-s integrations which are summed to form the LC and
SC data on-board. In this work we used only long-cadence data, which in most of the cases are sufficient to discover and characterize the light curve features of RR\,Lyrae stars.

Due to bandwidth limitations, only pre-selected targets were observed. In the $Kepler$ field-of-view approximately 200,000 stars were observed, and on average 12-20 pixels ('apertures') were allocated per target. Photometric time-series data collected from only 5-6 percent of all pixels were downloaded, but these apertures contained a 1-2 pixel wide halo around the main targets. In addition, large, 200 x 200 pixel, contiguous areas ('superstamps') were also downloaded, primarily centered on open clusters NGC\,6791 and NGC\,6819 \citep{kuehn2015}. 
The superstamps and the generous extra pixels of the main targets allow us to find faint, background objects in the main target apertures.

\textit{Kepler} also observed large background fields on each CCD modules 
('EXBA' or EXtended BAckground sources/masks) starting from Q5 till the end of the primary {\it Kepler} Mission, for their first scientific exploitation see \citet{Martinez-Palomera2021}. We did not used EXBA pixels in this work.

\subsection{Pixel-level manipulation}

We downloaded all \textit{Kepler} Q4 LC target pixel files (TPFs) from \textit{Kepler}'s original mission including superstamps. We did not use SC data, but it is worth noting that for targets observed with the short (1-min) cadence, LC observations are also available, and given the typical timescales of RR\,Lyrae light variations LC data suffice. Individual pixels were generated from TPFs using the software package {\tt fitsh} \citep{2012MNRAS.421.1825P}, and pixels belonging to a given main target were handled together. 

We chose Q4 as the reference quarter in the Kepler Pixel Project, because by that time the \textit{Kepler} mission got rid of initial 'teething' problems, and Q4 became the first particularly quiet quarter. For example, guiding stars with an assumed constant flux are used to maintain the  constant attitude of the spacecraft. If variable stars used instead, the algorithm that ensures the correct pointing can fail.
Most problematic variable guiding stars were replaced in Q2, the centroiding algorithm was updated in Q3. In addition, much less attitude tweaks were needed during Q4 than previously, and no loss of fine point events were experienced   \citep{KDCH}.
The failure of one of the CCD-modules in Q4 caused gaps in the light curves of objects falling on the bad modules. In addition, small gaps are seen in the light curves obtained in Q4 and all other quarters. These are due to unplanned safe modes and loss of fine point events, as well as regular data downlink periods.

For the RR\,Lyrae stars this means that only Q4 LC data were searched for RR\,Lyrae like variations, but after finding them, we analyzed all data from Q0 to Q17 for a given candidate. The pointing of Kepler spacecraft was good enough to re-point itself well within 1 pixel accuracy between download periods, but the target list and also the size of the allocated apertures changed slightly from quarter to quarter. Therefore in our approach a few RR\,Lyrae candidates may have been lost if they fall on the apertures of a  target not observed in Q4 or were cut off because of the optimization of the size of the aperture masks.

We analyzed all pixels individually, irrespective of  whether we had known variable stars, non-variable main targets or background pixels. The total number of analyzed pixels was 4,949,464. 

\subsection{Search for RR Lyrae stars} \label{sec:varsearch}

To find RR\,Lyrae stars we computed the Lomb-Scargle periodogram of each individual pixel light curve 
using the astroML Python module\footnote{https://www.astroml.org/} \citep{astroML}. All the periodograms were analyzed by a Fortran code automatically. 
We flagged RR\,Lyrae candidates using the following criteria:
\begin{enumerate}
\item the main period had to fall between 0.2 and 1.0 days 
\item besides the main peak, the algorithm had to find at least two harmonics of the main frequency with decreasing amplitudes. 
\end{enumerate}

The algorithm returned 16830 individual pixels based on these criteria. We employed visual inspection to retain genuine RR Lyrae stars. This was necessary because the vast majority of our candidates were in fact eclipsing binary systems. These systems will be the main subject of other papers in this series (\cite{bienias2021} and Forró et al., in prep.)

The search yielded 40 RR Lyrae candidates in total, and in the majority of the cases multiple pixels were identified that belonged to each candidate. We analyzed the TPFs containing the candidates to visually determine and carefully select the optimal aperture for photometry, in a way of making sure to add all pixels showing the candidate signal, while minimizing contamination from the main target at the same time. In this step we used the {\tt Lightkurve} package \citep{Lightkurve}. An example is shown in Fig.~\ref{fig:pix_sel}. 
The selection was particularly challenging in cases where the candidate was located close to the main target or where the returned pixel signal was quite faint or noisy. We located the candidates in other quarters' TPFs by their celestial coordinates, which we derived from their Q4 pixel coordinates. The corresponding apertures were determined in the same manner as their Q4 apertures were, and finally the LCs from all available quarters were stitched. For most candidates, we could produce a continuous 4-year-long light curve. In case of certain candidates there are quarters where either the specific CCD module was not making observations or the candidate simply fell outside of the downloaded aperture.

We checked Table\,13 of \citet{vancleve2016}
to find those targets that fall on  'noisier' channels in any quarters. 
Nine candidates are free from any problems, most RR\,Lyrae candidates have minor issues (mild Moiré pattern, undershoot, elevated read noise) in some quarters. Three targets had severe Moiré and 'rolling band' patterns \citep{vancleve2016} in selected quarters (KIC\,5953307, KIC\,7440490, KIC\,9714117) according to the table, but their light curves are not affected. 

Reassuringly, our method flagged all known RR\,Lyrae variable stars in the Kepler field (\citet{benko2010, nemec2011,moskalik2015}).

\subsection{Cross-match of the RR Lyrae candidates}

We cross-matched our candidates with the Gaia EDR3 catalog \citep{gaiaedr3-2021}. The Gaia catalog was searched for stars located in the 20" radius of the coordinates of the candidates, which was necessary because in many cases the TPFs contain only partial images of the candidates, so those pixel coordinates do not always refer to the center of the Point Spread Functions (PSF). All possible matches were analyzed and the correct match was selected manually. For three of our candidates we found no Gaia match, these were removed from the candidate list (see Sec~\ref{sec:vetting} for details). Then we cross-matched the candidates with the Gaia DR2 RR Lyrae catalog of \citet{clementini2019}, in which 19 of our candidates were listed. 
All our candidates that have a counterpart in the Gaia RR Lyrae catalog show consistent periods derived from the two space missions' data. We found matching Pan-STARRS objects in the case of seventeen candidate RR Lyrae stars \citep{SHM2017}. 

Interestingly, ten of our candidate RR Lyrae stars were independently found on the \textit{Kepler} Full-Frame Images by \citet{molnar-2018}. All these ten candidates are known RR Lyrae stars previously found by either Gaia or the Pan-STARRS survey.

\subsection{Candidates and vetting}\label{sec:vetting}

We started with 40 RR\, Lyrae candidates. During the vetting procedure we eliminated 11 of them based on a detailed analysis of their light curves and their stellar neighbourhood. We found that 6 of these candidates were actually other type of variable stars, eclipsing binaries or rotational variables. For the remaining 5 candidates being dropped from the list, no corresponding Gaia EDR3 stars were found during the cross-match and these pixels had light curves with poor signal-to-noise ratios. In the case of additional 3 candidates we decided to discard them, because of extremely low signal-to-noise ratios of their light curves. We ended up with 26 objects, and we consider these as firmly confirmed RR\,Lyrae stars. Five of them are new discoveries, see Sec.~\ref{sec:results}.

\subsection{Detrending and post-processing}

The raw light curves have been extracted by simple pixel-aperture time-series photometry, thus they are exposed to all kinds of systematics that are specific to the \textit{Kepler} instrument. Before any scientific analysis, we need to eliminate or at least minimise the effects of these spurious signals that are mainly caused by differential velocity aberration, thermal gradients across the CCD, and pointing variations. There are a wealth of existing pipelines and methods to remove systematic artefacts such as Pre-search Data Conditioning (PDC; \citealt{PDC}) and Cotrending Basis Vectors (CBVs; \citealt{CBV}). However, these methods provide high quality, systematics-free light curves for the main \textit{Kepler} targets only. Applying them to our tailor-made aperture photometry does not yield the desired result, so we decided to develop our own methods to correct the raw light curves. We came up with two approaches to solve this issue.

Before any systematic corrections, we detected and removed outliers using a two-step process. First, we used an iterative sigma clipping to remove points that are above or below the mean light curve by $6\sigma$ or $4.5\sigma$, respectively. The chosen $\sigma$ values prevented us from cutting the sharp extremes, the maxima of RR Lyrae light curves, but left the intermediate points untouched. Because of this, we then fitted a series of summed sines and cosines using the \texttt{w{\={o}}tan} package \citep{wotan}, with which we divided the light curve, and an iterative sigma clipping with $5.5\sigma$ and $10\sigma$ thresholds was applied again.

The first detrending method is similar to the normal \textit{Kepler} CBV correction technique, where the CBVs were constructed using a set of carefully selected quiet stars. Instead of this, we extracted photometry from the pixels that are not part of our tailor-made apertures. Then, these light curves were used to define different CBVs for each quarter. We used principal component analysis (PCA) that is implemented in the \texttt{Lightkurve} package to extract the most common variance (i.e., the systematic) across all the pixel-level light curves. We determined the number of PCA vectors that are used to build the CBVs separately for each quarter. The decisions were based both on maximisation of Combined Differential Photometric Precision (CDPP) of the corrected light curves and visual inspections. Due to the low flux level of non-target pixels, the raw vectors would introduce additional noise components. To correct this effect, we fitted each vector with a spline and saved the analytical fits instead of the raw results. During the correction step, the CBVs are fitted to, then removed from the extracted SAP light curves. The CBVs defined by the \textit{Kepler} team are provided not only for each quarter but for each CCD module as well. We compared our own CBVs for different modules and found that they are very similar so taking more time to get per-module vectors would not give us better results.

In some cases the CBV-corrected light curves still contained spurious variations probably originating from contamination of the main target or nearby brighter stars. As the source of the contamination is usually unknown and different for each target, we simply fitted and removed a spline from the raw light curves to eliminate these variations. Before the fit we determined the main pulsation period then used \texttt{w{\={o}}tan} to iteratively fit a spline with a window width of 2 or 5 times the period. The latter (i.e. the smoother) was applied if the mean flux level was below 200 e$^-$/s, i.e., in case of very noise data sets.

The \textit{Kepler} telescope was rolled by 90 degrees after each quarter, thus stars fell onto different CCD modules with slightly different sensitivities quarterly, which resulted in different mean flux levels (i.e. jumps) between the consecutive light curve sections. Because of this we needed to perform two additional light curve preparation steps, the stitching and normalisation of the quarters. Before the former step, we flagged the remaining outlier points using a percentile-based detection method not to distort the results. Then, we selected a given portion of the beginning and end of each consecutive quarter and scaled their mean flux levels and peak-to-peak variations to match. Usually we selected 10-pulsation-period-long cuts or the whole quarters. The decisions were made by visual inspections to minimise the effect of amplitude attenuation, which appears in the former case if there is a trend in the light curve. The final data set was normalised to the quarter which has the largest amplitude, as we assumed that this is the case where we lost the least amount of charge due to the fact that in each quarter a different portion of the whole PSF fell within the aperture.

\subsection{Period determination}

In order to search for RR Lyrae stars, Lomb-Scargle periodograms were computed for all individual pixels of each main target star's aperture in the original \textit{Kepler} field. This method was a robust and efficient way to yield the initial period which served as a basis for candidate identification (see Sec.~\ref{sec:varsearch} for details). Other, more computationally intensive period determination methods were then applied to the already existing candidate list, namely Phase Dispersion Minimization (PDM) and Fourier analysis. Comparing the periods calculated using each method, we found that they were in agreement, there were only subtle differences, typically in the $4^{th}$ or $5^{th}$ decimal digit. For consistency we use the Fourier method to calculate the final period for all objects. Fourier parameters were also calculated and used later (see in Sec.~\ref{sec:results}).

\section{Results} \label{sec:results}

As we mentioned in Sec.~\ref{sec:vetting} we ended up with 26 RR\,Lyrae stars. Table~\ref{tab:table1} contains coordinates and \textit{Kepler} IDs of the apertures in which these objects were found.\footnote{There are no KIC IDs associated with our new RR Lyrae stars, the KIC IDs refer to the main targets throughout this paper.} 
In addition, Gaia EDR3 and Pan-STARRS IDs are also provided, where available. In this work we assign a Kepler Pixel Project (KPP) ID to avoid confusion. In the first column we list the KPP IDs. We use these designations throughout this paper to refer to our objects.

Table~\ref{tab:table2} contains Gaia G and \textit{Kepler} Kp magnitudes, 
period(s), Fourier parameters, and RR Lyrae sub-type classification of our objects.

Out of the 26 confirmed RR Lyrae stars in our sample (these are found independently by our group), five had not been identified as RR Lyrae stars in neither Pan-STARRS nor Gaia, nor in any other catalogs. These are new discoveries. These four stars are KPP\#3, 12, 13, 15, 21. 

\subsection{Individual candidates}

\begin{figure*}
    \centering
    \includegraphics[scale=0.45]{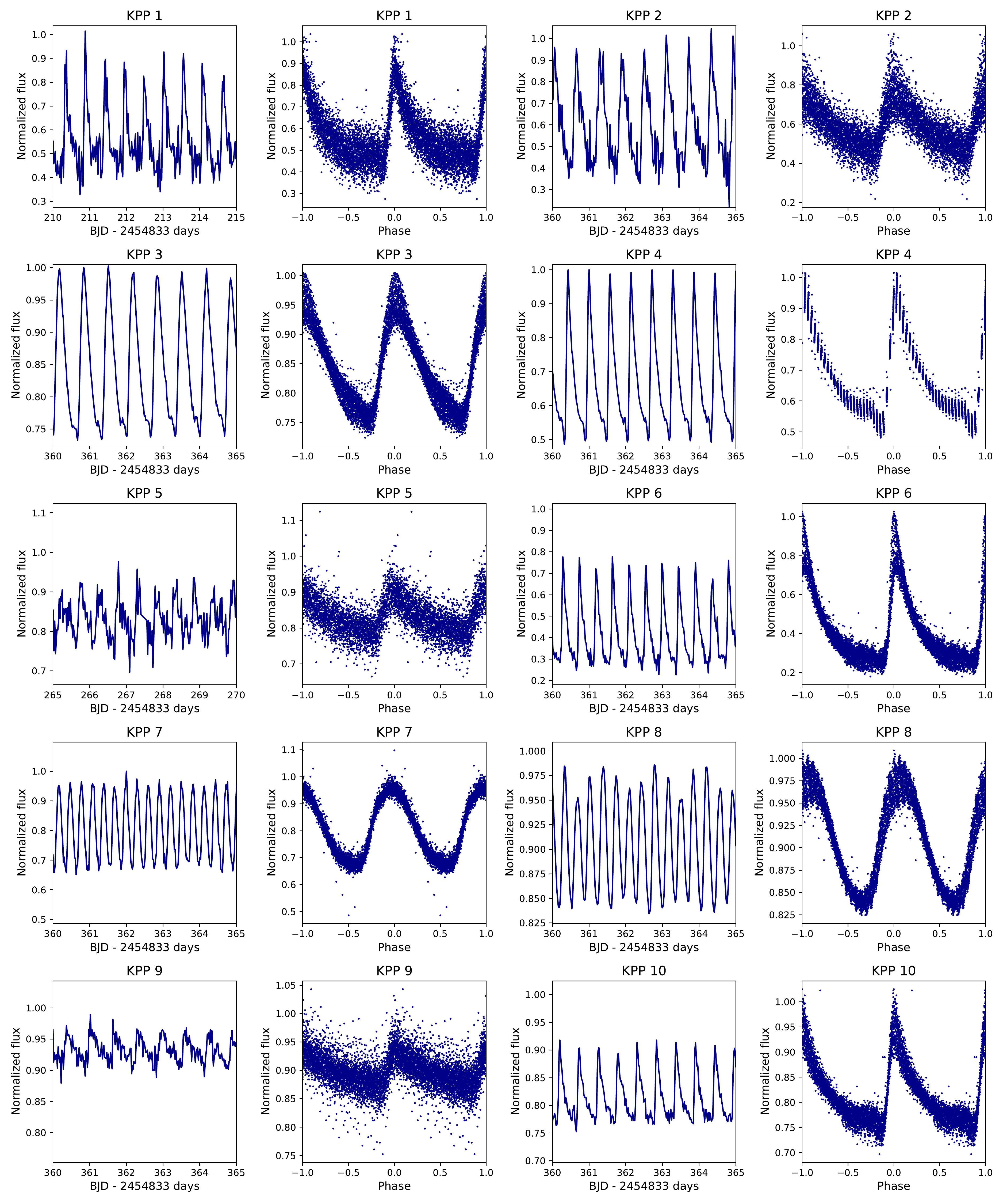}
    \caption{Five-day sections and phase-folded light curves of each candidate RR~Lyrae.}
    \label{fig:mosaic1}
\end{figure*}

\begin{figure*}
    \centering
    \includegraphics[scale=0.45]{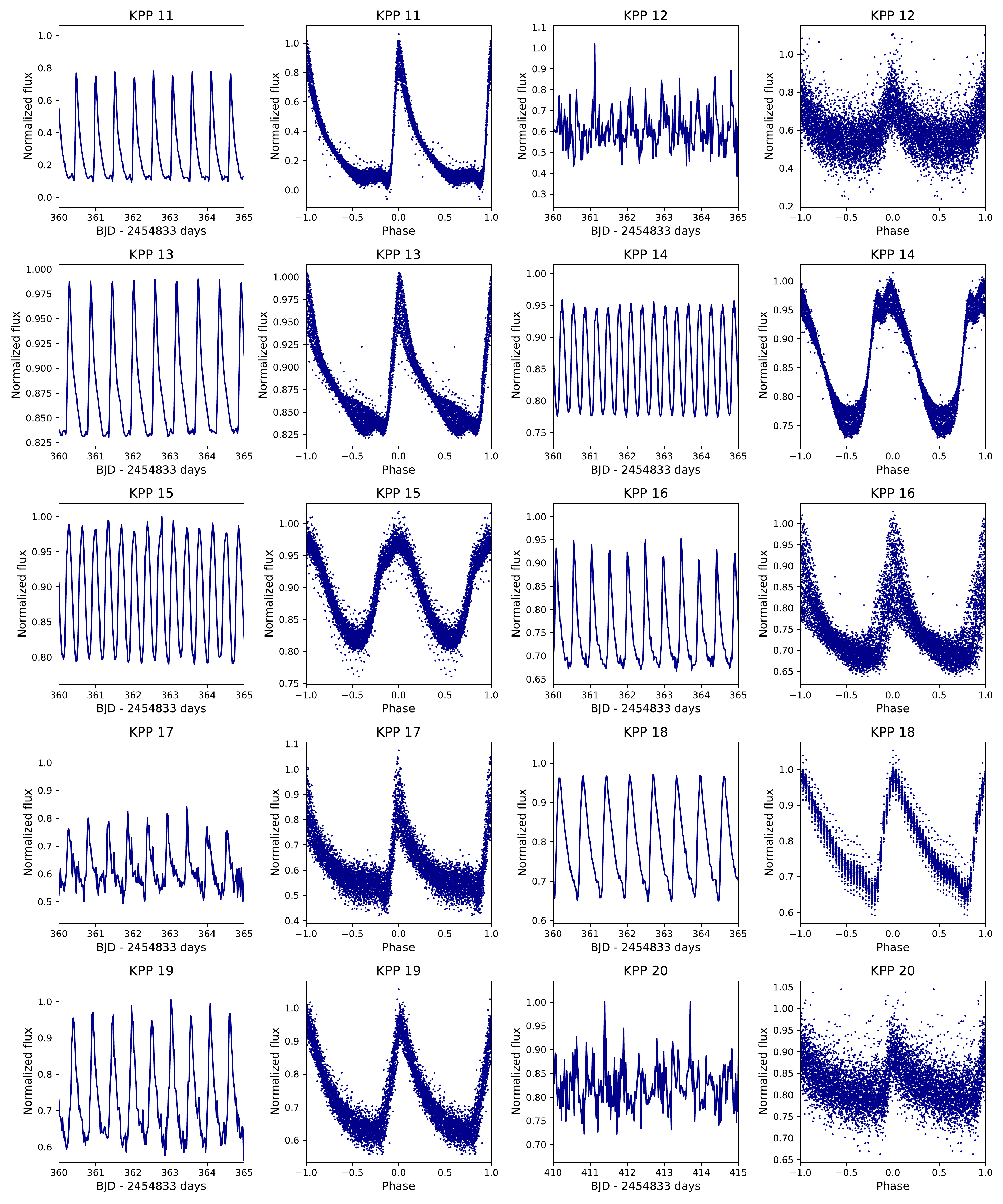}
    \caption{Fig~\ref{fig:mosaic1}. continued.}
    \label{fig:mosaic2}
\end{figure*}

\begin{figure*}
    \centering
    \includegraphics[scale=0.45]{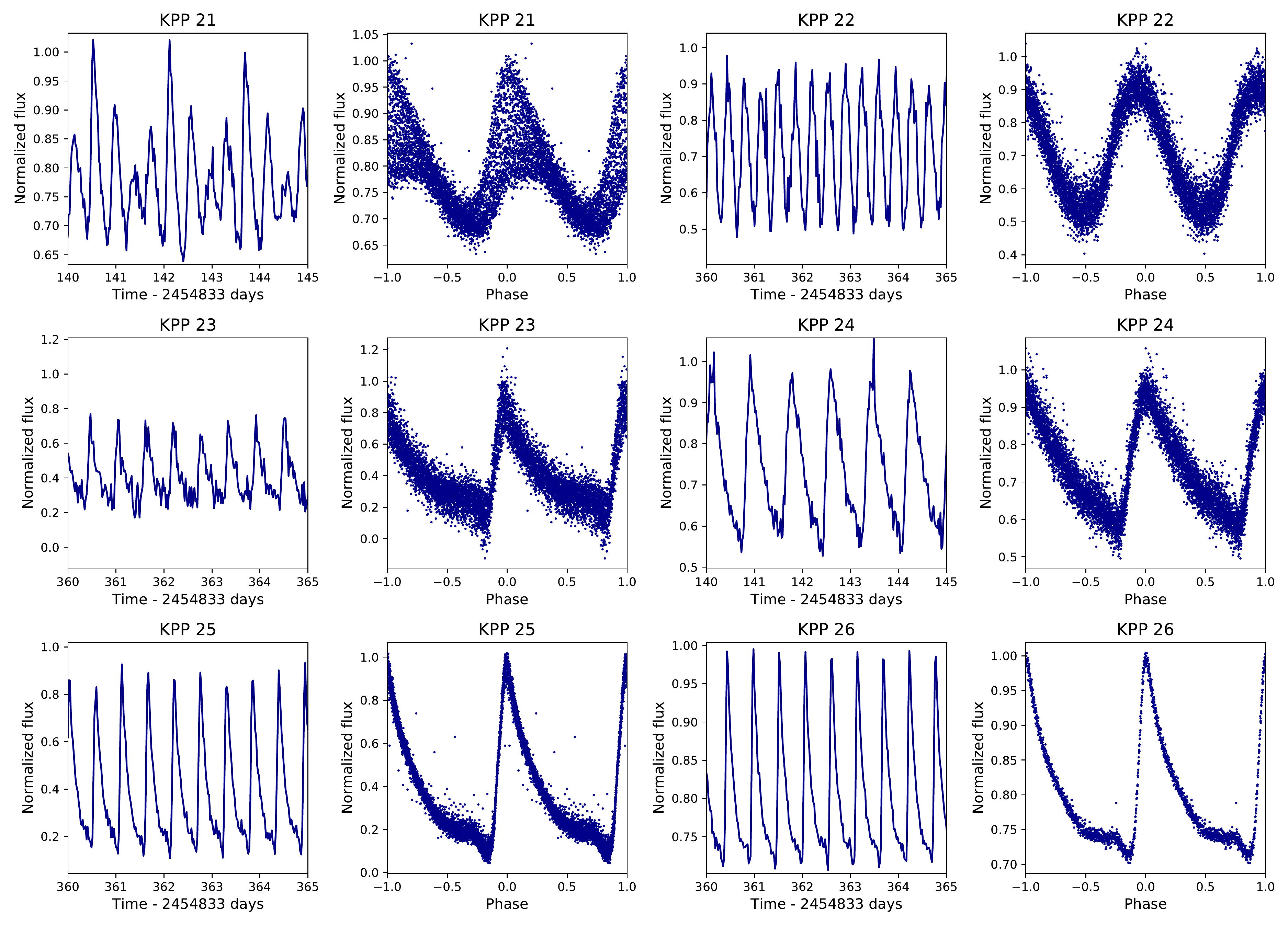}
    \caption{Fig~\ref{fig:mosaic1}. continued.}
    \label{fig:mosaic3}
\end{figure*}

\begin{deluxetable*}{cccccc}
\tablenum{1}
\tablecaption{Newly found RR\,Lyrae stars.  \label{tab:table1}}
\tablewidth{0pt}
\tablehead{
\colhead{KPP Nr.} & \colhead{KIC ID} & \colhead{Gaia EDR3 ID} & \colhead{R.A.} & \colhead{Dec.} & \colhead{Pan-STARRS IDs}
} 
\startdata
1 & 2449639 & 2051849058633285632 & 292.82119 & 37.69892 & J292.82122+37.69895\\
2 & 2831632 & 2099870400771768576* & 285.69669 & 38.07482 & J285.69670+38.07483\\
{\bf 3} & {\bf 3443012} & {\bf 2052744366034050048} & {\bf 291.28958} & {\bf 38.53009} & \nodata\\
4 & 3545038 & 2052740728204349952* & 291.54525 & 38.68922 & J291.54524+38.68924\\
5 & 4078693 & 2073225347082824320* & 296.58650 & 39.10820 & J296.58649+39.10821\\
6 & 5474028 & 2073781150208302592* & 298.25713 & 40.64793 &  \nodata\\
7 & 5953307 & 2101475172354523776* & 289.56251 & 41.28306 &  J289.56248+41.28309\\
8 & 5984662 & 2073877671026117248* & 298.05205 & 41.20253 & \nodata\\
9 & 6041330 & 2101562235632019712* & 291.46653 & 41.38119 &  J291.46653+41.38120\\
10 & 6049886 & 2077400197036707584* & 294.07810 & 41.30555 &  J294.07812+41.30556\\
11 & 6344281 & 2104492953811421696* & 283.85055 & 41.71403 &  J283.85055+41.71407\\
{\bf12} & {\bf6531416} & {\bf2077533508520642304} & {\bf294.56671} & {\bf41.971673} &  \nodata\\
{\bf 13} & {\bf 7115228} & {\bf 2101970639774939264} & {\bf 290.97815} & {\bf 42.60217} &  \nodata\\
14 & 7422855 & 2105082292044513536* & 283.58848 & 43.09951 &  \nodata\\
{\bf 15} & {\bf 7440490} & {\bf 2102040660629025280} & {\bf 290.43819} & {\bf 43.02845} &  \nodata\\
16 & 7869837 & 2105275698713637760* & 281.91247 & 43.69644 & \nodata\\
17 & 7904375 & 2079540980532315392* & 296.14073 & 43.61978 &    J296.14074+43.61979\\
18 & 8324032 & 2079035926741537664* & 299.14995 & 44.24873 &    J299.14995+44.24874\\
19 & 8824898 & 2079913573232819200* & 294.16275 & 45.04917 &    J294.16273+45.04921\\
20 & 9714117 & 2128087030238775040* & 293.43524 & 46.43196  &  J293.43526+46.43199\\
{\bf 21} & {\bf 9972328} & {\bf 2085608635442345984} & {\bf 299.27266} & {\bf 46.88490}  & \nodata\\
22 & 10141938 & 2128369501647067648 & 292.1582 & 47.16369 &  J292.15517+47.16122\\
23 & 10682408 & 2086510505563258496* & 297.76430 & 47.97088 &  J297.76436+47.97090\\
24 & 11038223 & 2134721101140387200* & 295.64509 & 48.56171 & \nodata\\
25 & 11229102 & 2143761530560524416* & 282.94103 & 48.95145 &  J282.94100+48.95145\\
26 & 11391636 & 2132005169981443200* & 285.28406 & 49.24156 &  J285.28405+49.24156\\
\enddata
\tablecomments{The first column contains the Kepler Pixel Project ID that we assigned to our objects. The second column contains the KIC ID of the main target of the aperture in which the variable candidate was found. In the third column an extra '*' means that the star was designated as an RR Lyrae variable by \cite{clementini2019}. In the last column PanSTARRS IDs are given \citep{SHM2017}. Boldface indicates new discoveries.} 
\end{deluxetable*}

\begin{deluxetable*}{ccccccccc}
\tablenum{2}
\tablecaption{Properties of our candidate RR\,Lyrae stars. \label{tab:table2}}
\tablewidth{0pt}
\tablehead{
\colhead{KPP Nr.} & \colhead{Gaia G} & \colhead{Kp} & \colhead{Period} & \colhead{R21} & \colhead{R31} & \colhead{Phi21} & \colhead{Phi31} & \colhead{Type}
} 
\startdata
1  &  16.34  &  19.28  &  0.53470502(7)  &  0.543(1)  &  0.384(1)  &  5.277(5)  &  4.542(7) & RRab+BL\\
2  &  18.09  &  19.59  &  0.6120445(1)  &  0.434(1)  &  0.229(1)  &  5.758(6)  &  5.468(9) & RRab+BL\\
{\bf 3}  &  {\bf 17.39}  &  {\bf 18.12}  &  {\bf 0.66834031(4)}  &  {\bf 0.3175(3)}  &  {\bf 0.1146(3)}  &  {\bf 5.796(2)}  &  {\bf 5.729(3)} & {\bf RRab+BL}\\
4  &  16.86  &  18.79  &  0.57217677(4)  &  0.5298(6)  &  0.3534(6)  &  5.495(2)  &  5.017(4) & RRab\\
5  &  17.18  &  20.58  &  0.5172499(2)  &  0.467(3)  &  0.240(3)  &  5.66(1)  &  5.34(2) & RRab\\
6  &  19.59  &  20.07  &  0.45121696(5)  &  0.540(1)  &  0.349(1)  &  5.327(5)  &  4.536(7) & RRab+BL\\
7  &  17.35  &  20.25  &  0.30411602(2)  &  0.1280(8)  &  0.0969(7)  &  5.595(8)  &  0.06(1) & RRc\\
8  &  17.72  &  17.30  &  0.35267846(2)  &  0.1077(4)  &  0.0601(4)  &  6.155(4)  &  0.430(6) & RRd\\
  &    &    &  0.4744649(9)  &  0.047(9)  &  0.008(8)  &  5.3(2)  &  4(2) & \\
9  &  16.99  &  19.70  &  0.6422573(6)  &  0.558(7)  &  0.364(7)  &  5.59(3)  &  5.14(4) & RRab\\
10  &  17.53  &  18.99  &  0.52882158(7)  &  0.5088(9)  &  0.3762(9)  &  5.470(5)  &  4.846(7) & RRab\\
11  &  16.62  &  18.30  &  0.51973399(5)  &  0.5170(6)  &  0.3855(6)  &  5.280(4)  &  4.427(5) & RRab\\
{\bf 12}   &  {\bf 17.51}   &  {\bf 19.91}  & {\bf 0.4650911(1)}   & {\bf  0.413(2)}   & {\bf  0.208(2)}   &  {\bf 5.13(1)}   &  {\bf 4.01(2)}   & {\bf RRab+BL} \\
{\bf 13}  &  {\bf 18.21}  &  {\bf 15.88}  &  {\bf 0.57957077(4)}  &  {\bf 0.4863(6)}  &  {\bf 0.3150(5)}  &  {\bf 5.545(2)}  &  {\bf 4.900(3)} & {\bf RRab+BL}\\
14  &  15.80  &  16.65  &  0.30940258(1)  &  0.1554(2)  &  0.0791(2)  &  6.153(2)  &  6.201(3) & RRc\\
{\bf 15}  &  {\bf 17.97}  &  {\bf 17.56}  &  {\bf 0.3529061(2)}  &  {\bf 0.060(3)}  &  {\bf 0.026(3)}  &  {\bf 5.85(6)}  &  {\bf 5.3(1)} & {\bf RRc+BL}\\
16  &  17.72  &  18.46  &  0.4819532(2)  &  0.355(2)  &  0.142(1)  &  5.29(1)  &  4.34(2) & RRab+BL\\
17  &  16.60  &  18.99  &  0.53441403(6)  &  0.530(1)  &  0.3875(9)  &  5.358(4)  &  4.678(6) & RRab+BL\\
18  &  17.30  &  17.50  &  0.63349194(4)  &  0.4175(3)  &  0.2273(3)  &  5.758(2)  &  5.573(3) & RRab\\
19  &  17.12  &  19.70  &  0.52921345(5)  &  0.4209(7)  &  0.1891(7)  &  5.355(4)  &  4.514(6) & RRab+BL\\
20  &  15.71  &  21.52  &  0.45932(2)  &  0.5(5)  &  0.4(5)  &  5(2)  &  5(3) & RRab\\
{\bf 21}  &  {\bf 17.66}  &  {\bf 19.10}  &  {\bf 0.39551656(8)}  &  {\bf 0.205(1)}  &  {\bf 0.079(1)}  &  {\bf 6.27(1)}  &  {\bf 5.96(2)} & {\bf RRd}\\
 &    &   &  {\bf 0.5302328(3)}  &  {\bf 0.172(3)}  &  {\bf 0.034(3)}  &  {\bf 5.38(2)}  &  {\bf 5.17(9)} & \\
22  &  13.98  &  20.57  &  0.34873956(4)  &  0.0877(8)  &  0.0617(9)  &  6.12(1)  &  0.07(2)  & RRc\\
23  &  17.63  &  20.87  &  0.57620074(9)  &  0.515(2)  &  0.366(1)  &  5.716(5)  &  5.367(8) & RRab\\
24  &  19.10  &  20.97  &  0.8342683(1)  &  0.4044(7)  &  0.2078(6)  &  5.771(3)  &  5.581(5) & RRab\\
25  &  15.73  &  19.21  &  0.54594504(4)  &  0.5447(5)  &  0.3651(6)  &  5.361(2)  &  4.761(3) & RRab\\
26  &  17.90  &  17.85  &  0.54256417(5)  &  0.5436(7)  &  0.3878(7)  &  5.400(3)  &  4.799(4) & RRab\\
\enddata
\tablecomments{The first column contains the Kepler Pixel Project IDs. In the second column Gaia G magnitude is given. The third column contains Kepler Kp magnitudes. Columns 4-8 contain the period and the epoch-independent Fourier-parameters. In the last column we list the subtypes as a result of our classification. Boldface indicates new discoveries.}
\end{deluxetable*}

All light curves for the members of our final sample are provided in  Figs~\ref{fig:mosaic1}, ~\ref{fig:mosaic2}, and ~\ref{fig:mosaic3}. Two types of light curves are included. The left panels are five-day sections of the detrended, normalized light variation from Q4, except for a few cases where Q4 was of poor quality. In these cases we selected another suitable quarter for visualization. The right panels are the phase-folded light curves created with the best period derived from Fourier analysis. Some of the candidates that were found on only a few pixels show relatively large scatter. This is not surprising, since these are faint background objects and some of them are located near the edge of the aperture, therefore a fraction of the flux is lost.  In some cases modulation is clearly visible despite the high scatter. We performed Fourier analysis on all detrended light curves. 
Based on the light curves' shapes and the Fourier analysis we established the classification of all our candidates. 
We also compute period ratios of additional modes with low amplitudes that are now routinely found in high-precision light curves. We plot these on the Petersen-diagram (period – period-ratio diagram), which is an extremely useful diagnostic for mode identification and even for estimating stellar parameters \citep{kovacs2001}. Most notable frequency groupings are the so-called $f_{0.61}$ or $f_X$ periodicities, that are found in RRc and RRd stars showing a period ratio between 0.60 and 0.64 with the radial overtone pulsational mode. Most probable explanation is the presence of nonradial modes in these objects. 
It is worth mentioning that in some rare cases additional modes with period ratios ($P_{O1}/P_{X}$) around 0.68 can be found in RRc stars \citep{netzel2019}.
In the following we present the results derived for our objects by sub-classes.

\subsubsection{Non-modulated RRab stars} \label{sec:nBL}

Based on the light curves and the Fourier analysis we found 11 non-modulated RRab stars. These are KPP\#4, 5, 9, 10, 11, 18, 20, 23, 24, 25, 26 (see Table~\ref{tab:table1}). For these objects no modulation is seen in the phase-folded or individual light curves, and no additional frequencies above the  $4\sigma$ noise limit were found in the spectrum. We paid extra attention to the frequency interval between $f_0$ and $2f_0$, where $f_0$ is the frequency of the fundamental (longer period) radial mode, since this is the interval where the most prominent additional frequencies are expected.

There are three objects where peculiarities were found. In the case of KPP\#23 a tentative modulation of 61 days was detected, while in the light curve of KPP\#26 we see strong quarter-specific trends. All these variations are most probably of instrumental origin. 

In the case of KPP\#25---while no modulation is detected---we see an additional, very low amplitude\footnote{The amplitude of this periodicity is more than 200 times smaller than that of the dominant pulsation mode.} variation at $f=2.4655$ c/d with a period ratio of 0.7446 with respect to the fundamental mode, which means that this variation might be caused by first radial overtone pulsation. Interestingly in case of RR Lyrae, the eponym of its class, similar low amplitude variation was found which was attributed to the first radial overtone mode based on hydrodynamical model calculations \citep{molnar2012}. This period ratio – if indeed is a manifestation of a very asymmetric amplitude ratio double mode pulsation shown by RR Lyrae itself as well – places KPP\#25 right on the classical RRd ridge on the Petersen-diagram (see Fig~\ref{fig:petersen}).

\subsubsection{Modulated RRab stars}

 The candidates found in the apertures of the main targets KIC\,2449639 (KPP\#1), KIC\,2831632 (KPP\#2), and KIC\,3443012 (KPP\#3)
seemed to be non-modulated RRab stars. However, we found strong asymmetric triplet (sometimes quintuplet) features around the main frequencies and all of their harmonics. Therefore, we classify these objects as a Blazhko RRab stars. KPP\#1, KPP\#2, and KPP\#3 have modulation periods of 42.5\,$\pm$\,0.1 d, 40.5\,$\pm$\,0.1 d, 33.0\,$\pm$\,0.1 d, respectively.
In retrospect, in some of the cases modulation is indeed visible in some sections of the light curves as amplitude modulation.  We found no additional frequencies above the noise limit between $f_0$ and $2f_0$ in these objects.



 The RR Lyrae candidates found in the apertures of the main targets KIC\,5474028 (KPP\#6) and KIC\,7904375 (KPP\#17) seemed to be modulated RRab stars. In case of KPP\#12 in the aperture of KIC\,6531416 a strong modulation was seen already on the preliminary light curve.  We found strong asymmetric triplet (and sometimes quintuplet) features around the main frequencies and all of their harmonics suggesting  118\,$\pm$\,1 d,  31.9\,$\pm$\,0.2 d and 46 \,$\pm$\,1 d modulation periods in KPP\#6, KPP\#17, and KPP\#12, respectively. We found no additional significant frequencies above the noise limit between $f_0$ and $2f_0$.

The candidate found in the aperture of the main target KIC\,8824898 (KPP\#19) seems to be a modulated RRab star. We found strong asymmetric triplet features around the main frequency and all of its harmonics suggesting a 44.9\,$\pm$\,0.5 d period of modulation. 
In addition, we see frequencies characteristic of the period-doubling effect around 3/2 $f_0$. 

The candidate found in the aperture of the main target KIC\,7115228 (KPP\#13) seemed to be modulated RRab star. We found strong asymmetric triplet (and sometimes quintuplet) feature around the main frequency and all of its harmonics suggesting a 64\,$\pm$\,1 d period of modulation. We note that in Fig.~12 of \cite{molnar2022} there are additional modes at the fundamental mode period of 0.5796 at around 0.70 period ratio, however, between 0.62 and 0.64 the Petersen-diagram looks empty. However, in KPP\#13 we found two additional frequencies above the noise limit between $f_0$ and $2f_0$ at 2.4492 c/d, 2.7267 c/d resulting in 0.7045 and 0.6326 period ratios with the fundamental mode, respectively. This unique period ratio requires more investigation.

\begin{figure}
    \centering
    \includegraphics[scale=0.68]{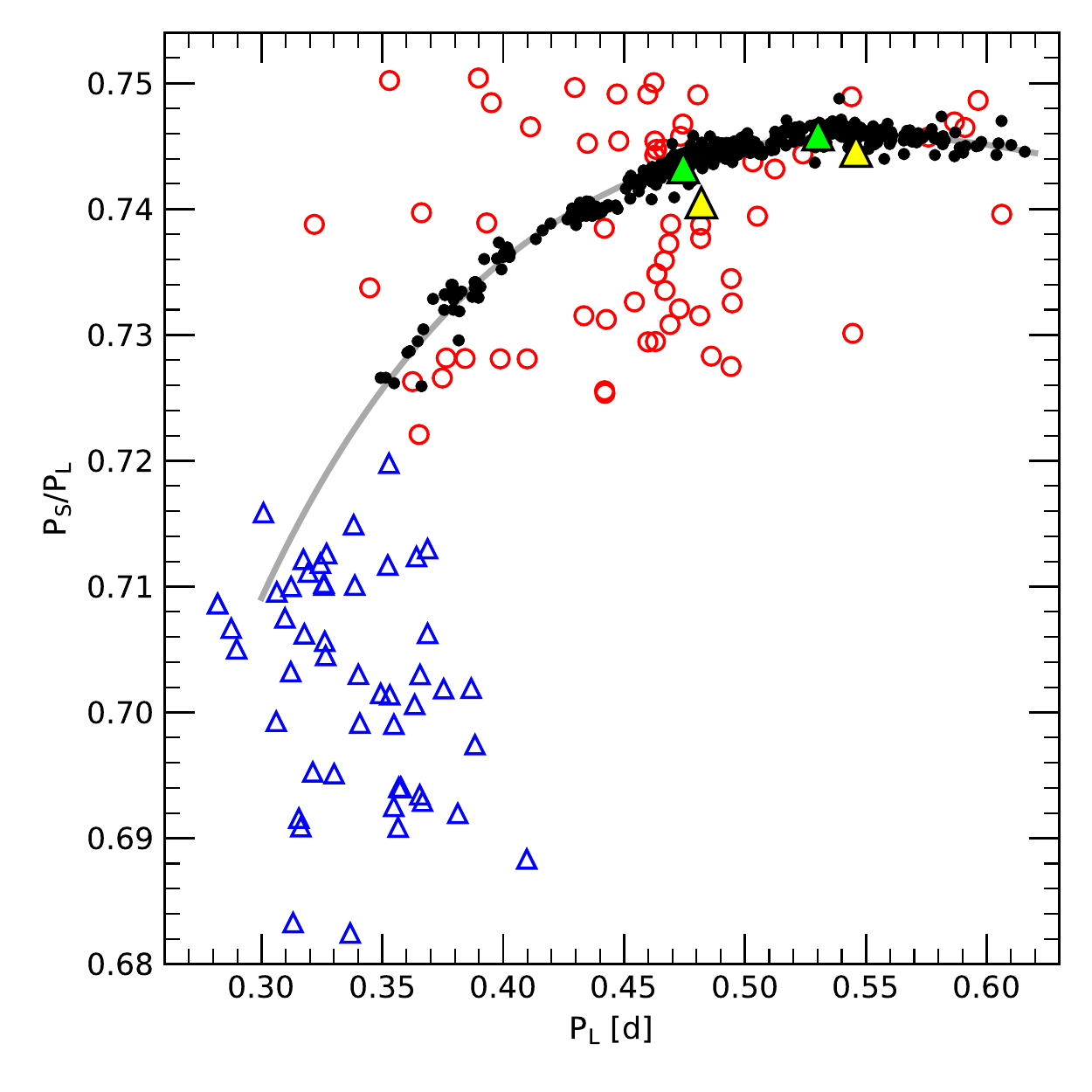}
    \caption{Petersen diagram: $P_{long}/P_{short}$ vs $P_{long}$. Black dots denote classical RRd stars  based on OGLE and K2 observations, red circles show the anomalous RRd (aRRd) stars, blue triangles show the peculiar RRd stars (pRRd) stars. Green triangles show our two new RRd stars, while the yellow triangles show KPP\#25, a non-modulated RRab star with an additional frequency that places it among the classical RRd stars, and KPP\#16, a modulated star with frequency content similar to RR Lyrae, the eponym of its class. 
    Figure is adapted from Fig~1. of \cite{nemec2021}.}
    \label{fig:petersen}
\end{figure}

The candidate found in the aperture of the main target KIC\,7869837 (KPP\#16) is a strongly modulated RRab star. We found strong asymmetric quintuplet (and sometimes beyond that) features around the main frequency and all of its harmonics suggesting a 60.6\,$\pm$\,0.6 d period of modulation. We found two additional frequencies above the noise limit between $f_0$ and $2f_0$ at 2.802 and 3.063 c/d. The first one gives a 
period ratio of 0.7405 which places it on the canonical RRd sequence on the Petersen diagram (Fig.~\ref{fig:petersen}), while the second additional frequency shows a period ratio of 0.6774, characteristic of the period doubling phenomenon discovered in \cite{szabo2010}. Taken together these frequencies and period ratios, this object closely resembles RR~Lyrae itself, where, beyond the Blazhko effect and the period doubling, first overtone radial pulsation mode was detected both in the \textit{Kepler} data and in hydro models as well \citep{molnar2012}. Interestingly, the amplitude ratio of the overtone and fundamental radial pulsation modes is an order of magnitude higher ($A_0/A_1$ =20.5) in case of KPP\#16 compared to RR~Lyrae, the prototype, while the frequency ratios put both stars slightly offset from the canonical RRd period ratio ridge, KPP\#16 being a bit lower, and RR~Lyrae itself a bit higher \cite{molnar2012}.

\subsubsection{RRc stars}

\begin{figure}
    \centering
    \includegraphics[scale=0.68]{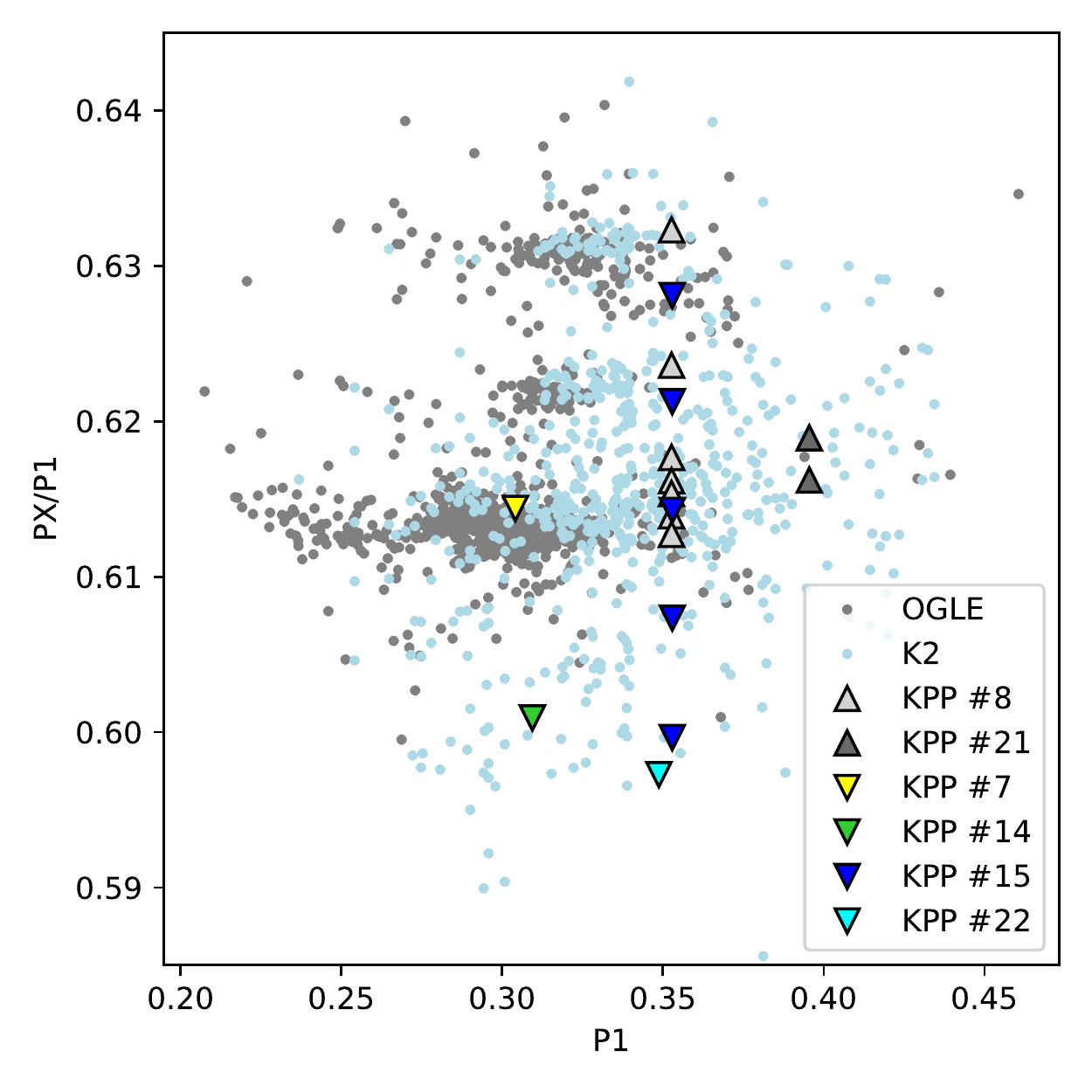}
    \caption{Petersen diagram focusing on the 0.61 period ratio characteristic of RR Lyrae stars pulsating in the first radial overtone mode (RRc and RRd stars).  Grey dots show the OGLE stars, light blue dots denote K2 objects, while color triangles correspond to our newly found RRc (downward triangles) and RRd (upward triangles) stars. Figure is adapted from Fig~3. of \citet{netzel2019} and unpublished K2 RR Lyrae data (Netzel et al., in prep.).}
    \label{fig:ridges}
\end{figure}

Fig~\ref{fig:ridges} shows the Petersen diagram focusing on the region of additional low-amplitude modes near the 0.61 period ratio in RRc stars. Large downward triangle symbols denote the frequency ratios of our newly found RRc stars against the background of small OGLE (grey) and K2 (light blue) dots. 

 The candidate found in the aperture of the main target KIC\,5953307 (KPP\#7) is an RRc star. We found a frequency at 5.351976 c/d which corresponds to the $P_{0.61}$ additional frequency with a 0.6143 period ratio, which places this star in the middle of the lower ridge of the additional modes commonly found in RRc stars \citep{netzel2019,molnar2022}.

Our next candidate found in the aperture of the main target KIC\,7422855 (KPP\#14) is also an RRc star. We found a frequency at 5.37835 c/d which corresponds to the $P_{0.61}$ additional frequency with a 0.6009 period ratio, which places this object below the main distribution of the lower ridge on Fig~\ref{fig:ridges}.

The candidate found in the background pixels within the aperture of the main target KIC\,7440490 (KPP\#15) is a phase-modulated RRc star featuring a characteristic double-maximum light curve. We found five frequencies with quasi-equidistant frequency separation at around 4.613 c/d which corresponds to the middle of the lower ridge  $P_{0.61}$ additional frequency family. The two frequencies with higher period ratios sit on the middle and upper ridges, respectively, while the two frequencies with the lower period ratios are located below the most populated lower ridge, but it is worth noting that these regions are also surprisingly well-populated despite the fact that currently no frequencies related to pulsational modes are predicted in this region.

Lastly, KPP\#22 in the aperture of KIC\,1014198 is a normal RRc featuring a 0.61 frequency at $f_X = 4.80100$ c/d.

We also note that no significant frequencies were found in any of our RRc stars around the expected 0.68 period ratio ($P_{O1}/P_{X}$) \citep{netzel2019}.

\subsubsection{RRd stars}

\begin{figure}
    \centering
    \includegraphics[scale=0.55]{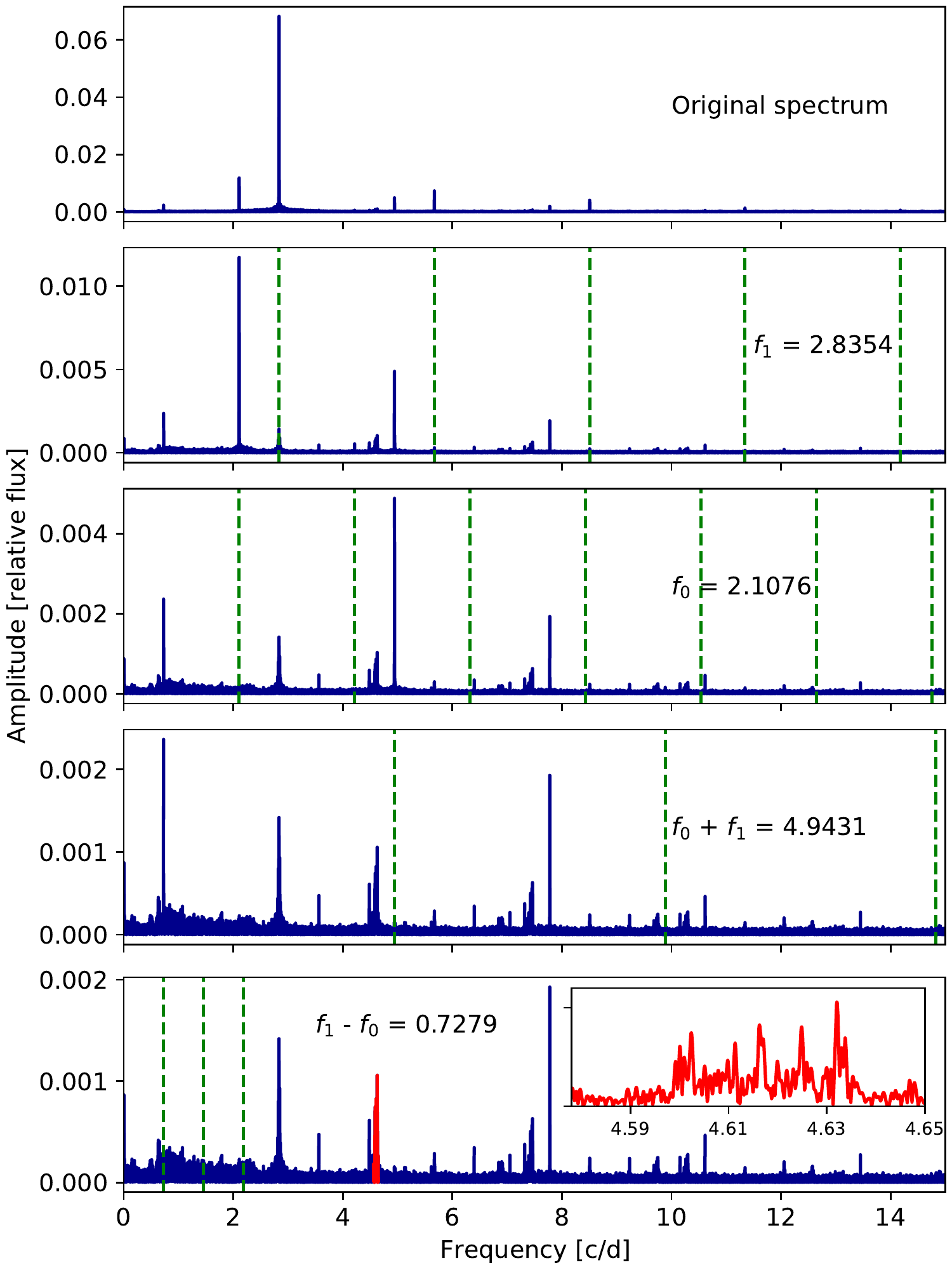}
    \caption{Fourier-spectrum of the newly found RRd star KPP\#8. The upper panel shows the original Fourier-spectrum. Then in each subsequent panel we subtracted the frequencies and their harmonics as shown by vertical green dashed lines. In the bottom panel we show in the insert the close-up of the $f_X$ ($f_{0.61}$ ) frequency structure.}
    \label{fig:spectr}
\end{figure}

The background RR\,Lyrae found in the aperture of the main target KIC\,5984662 (KPP\#8) is clearly an RRd star. Based on the  period of the first overtone mode (0.3527884\,d) and its period ratio ($P_S/P_L$= 0.7433) with the radial fundamental mode we classify it as a classical RRd star (cRRd) \citep{nemec2021}. 

Fig~\ref{fig:spectr} shows the Fourier-spectrum of KPP\#8 and a pre-whitening sequence with the first-overtone ($f_1$) and fundamental ($f_0$) modes and their linear combinations. Zooming in on the region of the $f_{0.61}$ frequencies, one can see a forest of closely packed frequencies. We see at least five peaks around the expected 0.613 frequency, the largest amplitude belongs to 0.6125 period ratio with the first overtone mode. This corresponds to the lower ridge in the Petersen-diagram of RRc stars showing the $f_{0.61}$ frequencies \citet{netzel2019}. The explanation of the existence of the forest of peaks around this frequency is most probably the non-stationarity of the pulsation mode. We see another frequency with f=4.484209 c/d (period ratio corresponding to 0.6323) that places this star on the upper ridge of the $f_{0.61}$ frequencies. Below our chosen detection limit ($5\sigma$) one can see extra power in the Fourier-spectrum in between the peaks belonging to the 0.61 and 0.63 ridges, hence we speculate that periodicity corresponding to the least populated middle-ridge is also present in this star, which makes this star a rare object among the classical RRd stars.

KPP\#21 is an RRd star, with a period ratio of 0.7459, the first overtone being the dominant frequency. With a fundamental mode period of 0.53023283\,d this is another example of a classical RRd star (cRRd) on the main sequence of canonical period ratio on the Petersen-diagram. The period of the first overtone mode is 0.39551656\,d. No other periodicities including the $f_{0.61}$ mode family were found in the Fourier spectrum. 

In Fig.~\ref{fig:petersen} we show the Petersen-diagram of a large number of classical (cRRd), anomalous (aRRd), and peculiar double-mode (pRRd) RR Lyrae stars \citep{arrd-2016,prudil-2017}. Our two newly discovered objects are denoted by large green star symbols. Based on their characteristics, these are classical, canonical RRd objects, while KPP\#25 
(denoted by a large yellow triangle and discussed in Sec~\ref{sec:nBL}) resembles the anomalous RRd class. 

Large upward triangle symbols denote the frequency ratios of the additional low-amplitude modes of our newly found RRd stars in the Petersen diagram in Fig~\ref{fig:ridges}. Similarly to our RRc stars no frequencies corresponding to the 0.68 period ratio were found in the discussed RRd stars.

\begin{figure*}
    \centering
    \includegraphics[width=\textwidth]{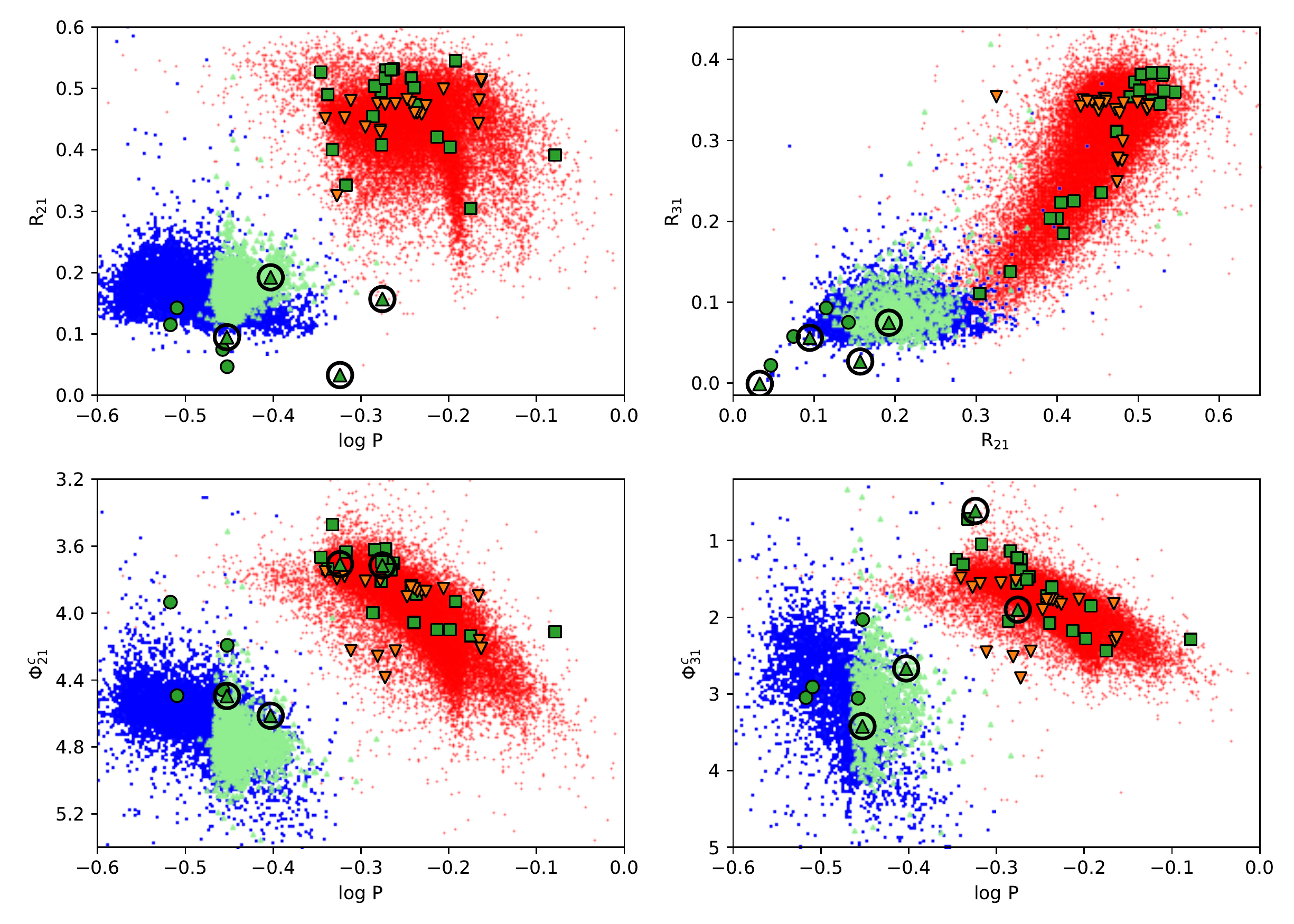}
    \caption{Comparison of Fourier parameters with OGLE-IV LMC field RR Lyrae stars from \citet{OGLE4RRL} and other known non-Blazhko Kepler RRab stars from \citet{nemec2011}.  The red points are RRab, blue squares are RRc, and light green triangles are RRd stars from OGLE. Downward orange triangles are known Kepler non-Blazhko RRab stars. Green squares, circles and triangles are the new RRab, RRc, and RRd stars, respectively, identified in this project. Both RRd stars are 
    shown by two green triangles in each panel. For better visibility we circled the markers representing the RRd stars.}
    \label{fig:fourpar}
\end{figure*}

 We performed standard  Fourier-analysis of the \textit{Kepler} photometry of all new RR\,Lyrae stars. Besides analyzing the Fourier spectra and its frequency content, it is interesting to compute the epoch-independent relative Fourier parameters as well. 
These parameters are the amplitude ratios of the Fourier-terms ($R_{i1}=A_i/A_1$, where $A_i$ is the amplitude of the $i^{th}$ Fourier-term), and the epoch-independent phase differences ($\phi_{i1}=\phi_i-i \phi_1$). These parameters describe the shape of the light curves, and correlate with important physical parameters of the RR\,Lyrae stars, such as metallicity \citep{jurcsik_kovacs1996}, and can also be used to classify our objects into RR\,Lyrae subclasses.
In Fig.~\ref{fig:fourpar} we show the first two amplitude ratios and phase difference parameters as a function of the logarithm of the dominant pulsation mode (except the upper right panel, where $R_{31}$ is plotted versus $R_{21}$). As a benchmark we plotted the Fourier parameters of the OGLE-IV LMC field RR\,Lyrae stars from \citet{OGLE4RRL} and other known non-Blazhko Kepler RRab stars from \citet{nemec2011}. The red points are RRab stars, blue squares are RRc stars and light green triangles correspond to the first radial overtone mode of the RRd stars from OGLE. Downward orange triangles are known Kepler non-Blazhko RRab stars. Green squares, circles and triangles are the new RRab, RRc and RRd stars, respectively, identified in this project. Both RRd stars are shown by two green triangles in each panel to represent their two radial pulsational modes. To account for the difference between the Kepler and OGLE photometric systems we used the linear relations found in \citet{nemec2011}. We conclude that the Fourier parameters allow an unambiguous identification of RR\,Lyrae stars confirming the classification of our objects.

\section{Discussion} \label{sec:discussion}

\begin{figure}
    \centering
    \includegraphics[width=8.5cm]{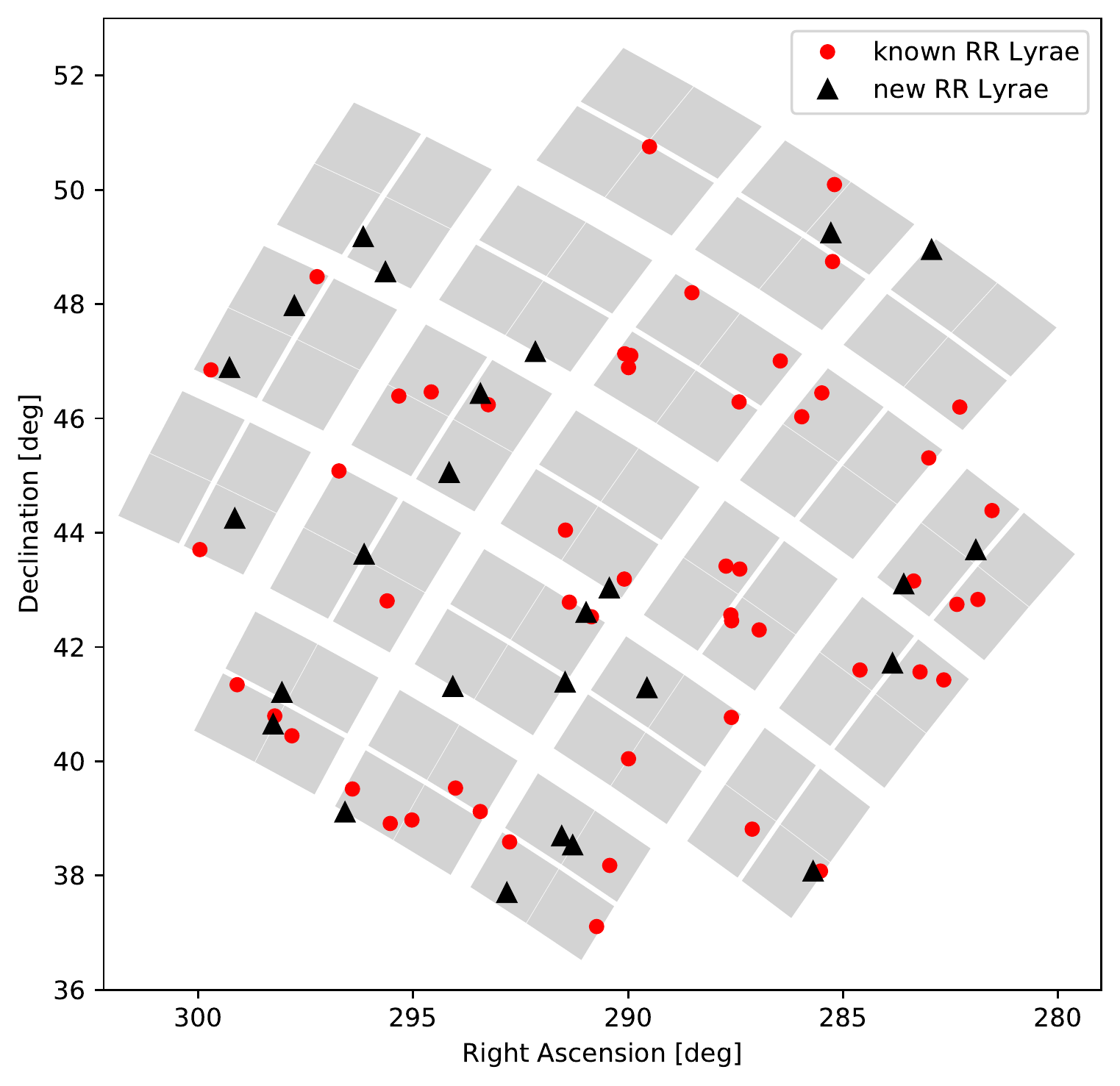}
    \caption{Known and new RR\,Lyrae stars in the original \textit{Kepler} field. Red dots show the already known RR Lyrae stars selected by the KASC, black triangles denote 26 new RR\,Lyrae stars found and confirmed by this project.}
    \label{fig:newrrl}
\end{figure}

Fig.~\ref{fig:newrrl} shows the  celestial coordinates of 52 RR\,Lyrae (RRab and RRc) stars that were known and selected before the start of the mission, or discovered by the KASC RR\,Lyrae and Cepheids Working Group (denoted by red dots) along with the FOV of individual CCD modules. The plot includes the extra 26 RR\,Lyrae stars found and/or confirmed by our work (shown by black triangles), which means a 50\% increase. Our sample includes RRab, RRc and also RRd stars with exquisite, 4-yrs long space photometry. The latter subcategory was missing from the original \textit{Kepler} sample.

\begin{figure}
    \centering
    \includegraphics[scale=0.65]{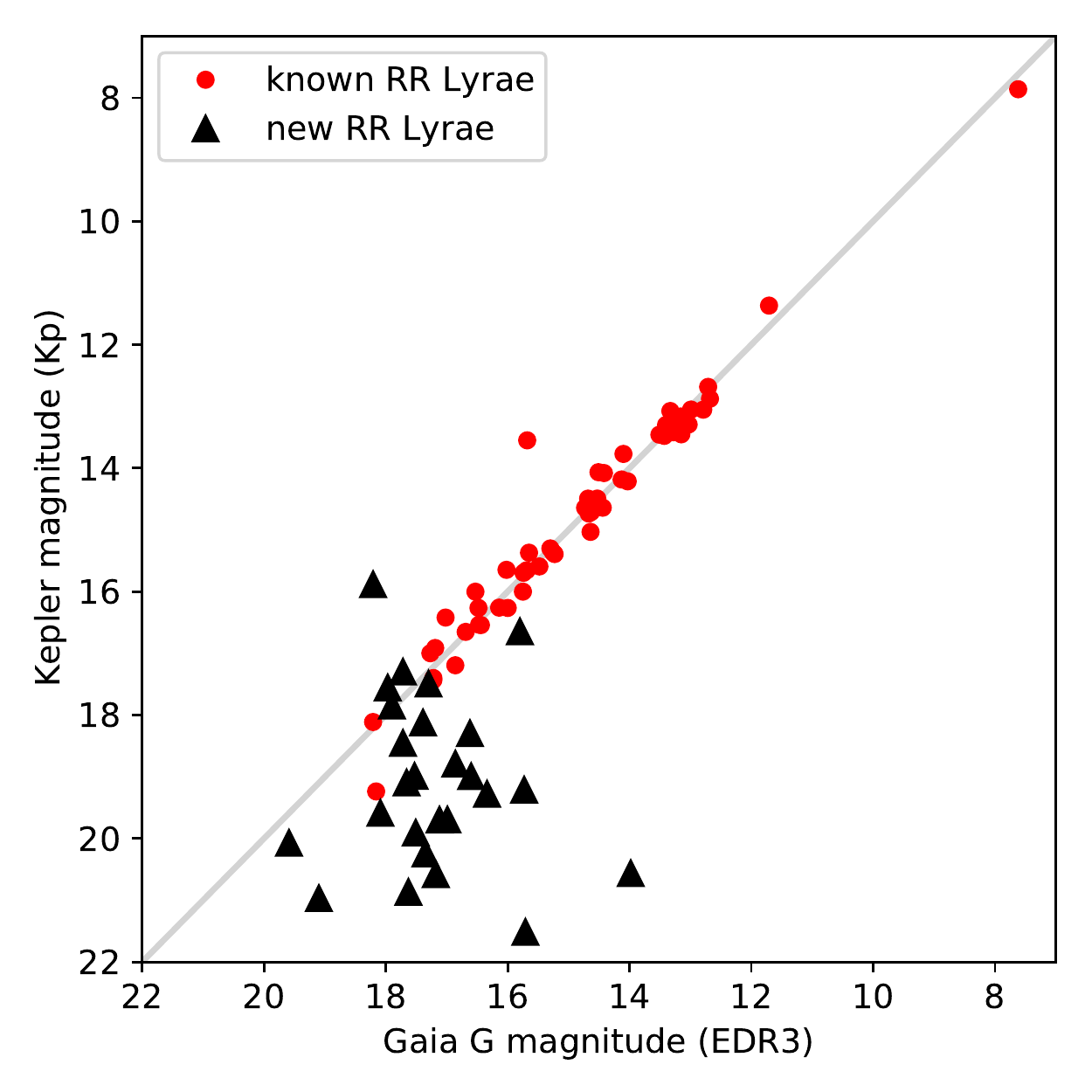}
    \caption{Gaia vs. \textit{Kepler} magnitudes of previously known (red dots) and newly discovered (black triangles) RR Lyrae stars. The 
    plot nicely demonstrates the amount of lost flux in the \textit{Kepler} apertures, since our targets in the new sample were not main targets of the \textit{Kepler} mission. Thus, orange points deviating downward significantly from the 1:1 line are faint 
    background objects, where some part of the flux was lost,  while stars contaminated by nearby bright stars are outliers above the 1:1 line.}
    \label{fig:G_Kp}
\end{figure}

In Fig.~\ref{fig:G_Kp} we demonstrate the consistency between Gaia and \textit{Kepler} magnitudes in case of previously known RR\,Lyrae stars (denoted with red symbols) and our new sample (black triangles). The plot nicely demonstrates the amount of lost flux in the main \textit{Kepler} apertures, since our targets in the new sample were not main targets of the \textit{Kepler} mission, and in some cases only a fraction of their flux was captured. Thus, black triangles deviating downward from the identity line, especially in the faint regime represent our background objects, where some part of the flux was lost. Among the known RR\,Lyrae stars, KIC\,7021124 (of Gaia magnitude 15.58, \textit{Kepler} magnitude of 13.55) is also an outlier, because it was contaminated by a nearby bright star in the Kepler observations. The newly discovered Blazhko RRab star (in the aperture of KIC\,7115228, KPP\#13, black triangle) deviating upwards from the identity line is contaminated by the main target in its aperture.
The brightest object at the top right corner of the diagram is RR\,Lyrae, the eponym of its class.

\begin{figure}
    \centering
    \includegraphics[scale=0.49]{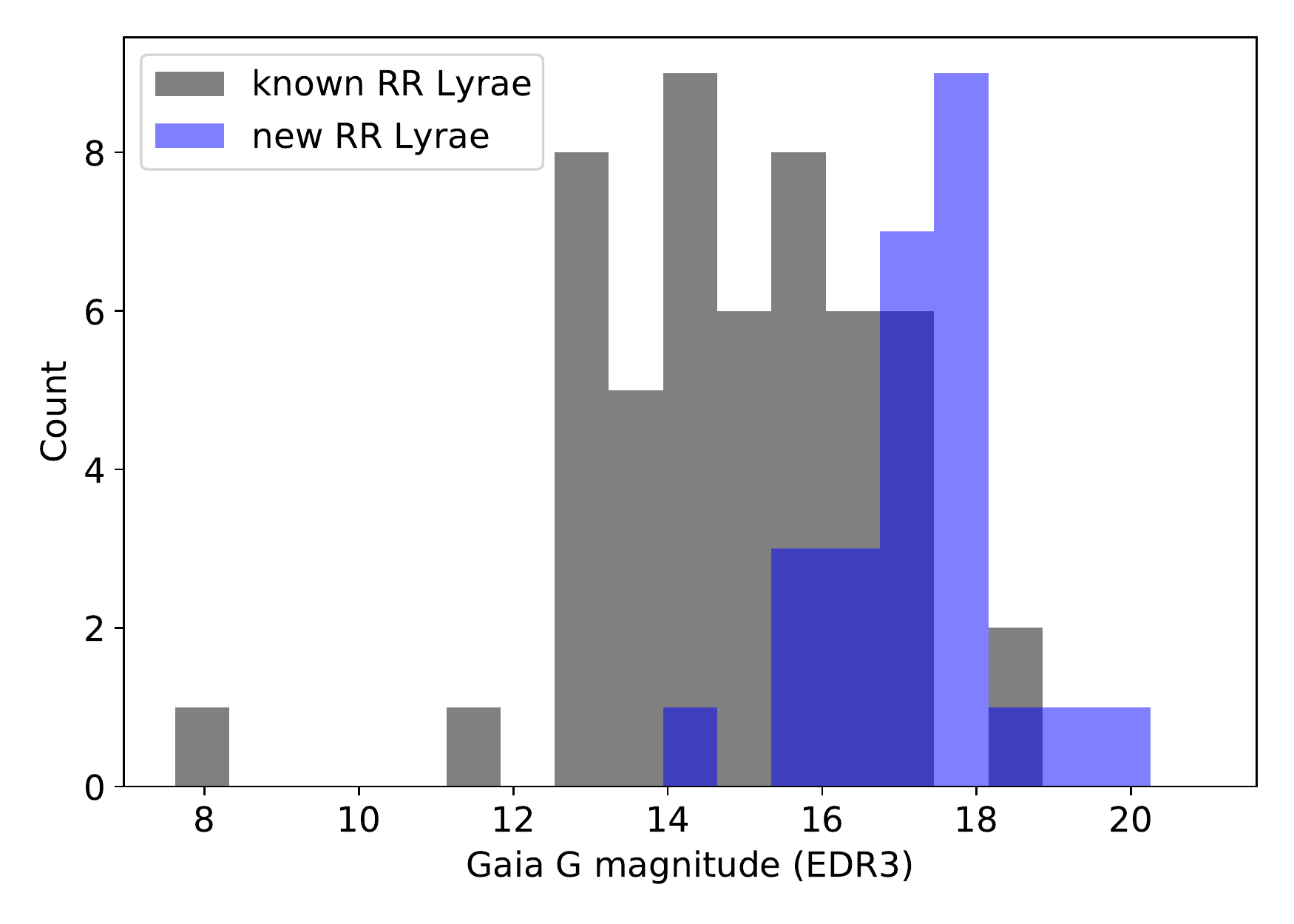}
    \caption{Gaia G magnitude distribution of the previously known (grey) and our new RR Lyrae sample (blue) in the original Kepler field.}
    \label{fig:gmag}
\end{figure}

In Fig.~\ref{fig:gmag} we plot the distribution of Gaia G magnitudes of the old and new RR\,Lyrae stars in the Kepler original field of view. It is not surprising that our background objects are at the faint end of the distribution. It is also conceivable that the majority of the new stars belong to the halo population (magnitude fainter than 16), while below 20th magnitude we look 'beyond' the Milky Way galaxy, therefore we do not expect to see many fainter objects in that direction. \textit{Kepler} was able to detect RR\,Lyrae stars fainter than 21 magnitudes during the K2 mission, but those belonged to a nearby dwarf galaxy \citep{molnar-2015}. The addition of the fainter end to the distribution is an important result of this work, especially considering the fact that despite the faintness of our objects, high-precision light curves are provided for each RR\,Lyrae star. 

During this work we found 20 RRab stars (9 of which show the Blazhko modulation), 4 RRc stars (one of them is Blazhko) and 2 RRd stars in the background pixels. The Blazhko occurrence rate is in accordance with previous studies: 45\% among RRab stars \citep{jurcsik2009} and a lower value for RRc stars (25\% in our case).

Regarding the frequency content of our RR\,Lyrae stars it is worth noting that in line with previous studies (especially based on space photometry) we found no extra frequencies in non-modulated RRab stars. Blazhko RRab stars show the characteristic frequency triplet or multiplet due to the modulation, and in some cases 3/2 $f_0$ frequency, as well, which is a sign of the presence of the period doubling phenomenon \citep{szabo2010}. Some other low-amplitude frequencies are present, but only in modulated RRab stars. Concerning RRc stars, all four of them show the $f_{0.61}$ frequencies. One of the RRd stars (KPP\#8) also shows this characteristic feature. 

Our objects nicely fit the Petersen-diagram of RRd stars (Fig.~\ref{fig:petersen}) and the one zooming in on the $f_{0.61}$ frequencies typically shown by RRc and RRd stars (Fig.~\ref{fig:ridges}). Various (and sometimes unexpected) parts of these diagrams are populated with more and more stars thanks to extended ground-based sky surveys and space photometric missions. Some of our RRc stars show more than one additional $f_{X}$ frequencies. These are especially important objects to test and understand the nature of these elusive, low-amplitude periodicities, which are currently hypothesized to be $l = 8,9$ nonradial modes and their linear combination   \citep{netzel2019, dziembowski2016}.

\section{Summary} \label{sec:concl}

In this work we continued our detailed investigation of the background pixels around the main targets and in superstamps observed in long-cadence mode during the original \textit{Kepler} mission. Fourier analysis revealed several pulsating stars, 26 of which are RR\,Lyrae-type variables. We confirmed their classification based on their 4-yr long light curves. Before this work 52 RR\,Lyrae stars were known in the original Kepler field. Thus, the 50\% increase in the number is a significant addition. 

Some of these objects were discovered by ESA's Gaia satellite and the Pan-STARRS survey independently, but the discovery of three new Blazhko RRab, one modulated RRc, and one RRd star are the results of the present work. 
Importantly, \textit{Kepler} provides high-precision photometric light curves containing thousands of data points in contrast with Gaia's considerably fewer photometric epochs. We emphasize that even in these faint objects (16--20 mag), found in background pixels with elevated noise levels and sometimes with significant flux loss, low-amplitude phenomena, such as the presence of additional periodicities could be investigated. The two new classical RRd stars also present a significant improvement, since so far no RRd stars were found in the \textit{Kepler} field. One of the two new RRd stars (KPP\#8) was previously classified as an RRc by the Gaia team. In addition, Blazhko-modulation was not known in any of these stars before this work. 

The modulation occurrence rates found among the new RRab and RRc stars are very similar to those derived from other ground-based and space-borne surveys. We see no significant difference in the occurrence of the Blazhko phenomenon in bright or faint RRab stars, thus those objects in the disk vs.\ the halo. This of course will have to be confirmed with more data. 

Importantly, all RRc (and one of the RRd) stars show the $f_{X}$ frequencies, i.e., low amplitude periodicities with 0.60--0.63 period ratios with the dominant radial first overtone pulsation mode. We note that if high precision (and long enough) space photometric data are available (e.g., CoRoT, Kepler) than these low-amplitude features seem to be ubiquitous in these sub-classes \citep{szabo2014, moskalik2015}. 

Background pixels have been found to be a treasure trove of time-dependent phenomena \citep{bienias2021}. This is due to the exquisite precision and long time span required and realized for exoplanet transit detections. We encourage the community to exploit these data in additional space-based photometric missions, as well, namely K2, TESS and the forthcoming PLATO space telescope.

Finally, we make the light curves of all the newly found and confirmed RR\,Lyrae stars publicly available at \url{https://www.konkoly.hu/KIK/data\_en.html}.

\acknowledgments

This project has been supported by the Lend\"ulet Program  of the Hungarian Academy of Sciences, project No. LP2018-7/2021, the KKP-137523 `SeismoLab' \'Elvonal grant of the Hungarian Research, Development and Innovation Office (NKFIH), and the EU's MW-Gaia COST Action (CA18104). 

This research made use of Lightkurve, a Python package for Kepler and TESS data analysis \citep{Lightkurve}. This paper includes data collected by the Kepler mission and obtained from the MAST data archive at the Space Telescope Science Institute (STScI). Funding for the Kepler mission is provided by the NASA Science Mission Directorate. STScI is operated by the Association of Universities for Research in Astronomy, Inc., under NASA contract NAS 5–26555. The authors gratefully acknowledge the entire Kepler team, whose outstanding efforts have made these results possible. 

This work has made use of data from the European Space Agency (ESA) mission {\it Gaia} (\url{https://www.cosmos.esa.int/gaia}), processed by the {\it Gaia} Data Processing and Analysis Consortium (DPAC,
\url{https://www.cosmos.esa.int/web/gaia/dpac/consortium}). Funding for the DPAC has been provided by national institutions, in particular the institutions participating in the {\it Gaia} Multilateral Agreement. 

The authors thank A. Pál for useful comments regarding the {\tt fitsh} package, Jim Nemec for helping with the Petersen-diagram (Fig.~\ref {fig:petersen}), H. Netzel for providing collected data of $f_{0.61}$ frequencies of OGLE and K2 RRc stars (Fig~\ref {fig:ridges}), and L. Moln\'ar for reading the manuscript and providing useful comments.

\vspace{5mm}
\facilities{Kepler, Gaia}

\software{Python \citep{Python3},
          AstroPy \citep{2013A&A...558A..33A},
          astroML \citep{astroML},
          LightKurve \citep{Lightkurve}, 
          fitsh \citep{2012MNRAS.421.1825P},
          W{\={o}}tan \citep{wotan}
          }

\pagebreak

\bibliography{kepler_pixel}{}
\bibliographystyle{aasjournal}

\end{document}